\documentclass[preprint]{aastex}

\usepackage{amsmath}

\shorttitle{Observations and Interpretation of a Low Coronal Shock Wave Observed in the EUV by the SDO/AIA } \shortauthors{Ma et al.}

\begin{document}


\title{Observations and Interpretation of a Low Coronal Shock Wave Observed in the EUV by the SDO/AIA}

\author{Suli Ma\altaffilmark{1,2}, John C. Raymond\altaffilmark{1}, Leon Golub\altaffilmark{1}, Jun Lin\altaffilmark{3},
Huadong Chen\altaffilmark{2}, \\
Paolo Grigis\altaffilmark{1}, Paola Testa\altaffilmark{1}, David Long\altaffilmark{1,4}}
\email{sma@head.cfa.harvard.edu} 

\altaffiltext{1}{Harvard-Smithsonian Center for Astrophysics, 60 Garden
        Street, Cambridge, MA 02138, USA}
\altaffiltext{2}{China University of Petroleum, 66 Changjiang West Road, Qingdao, Shandong 266555, China}
\altaffiltext{3}{Yunnan Astronomical Observatory, Chinese Academy of
Sciences (CAS), Kunming, Yunnan 650011, China}  
\altaffiltext{4}{Astrophysics Research Group, School of Physics, Trinity College Dublin, Dublin 2, Ireland}


\begin{abstract}
Taking advantage of both the high temporal and spatial resolution of the Atmospheric Imaging Assembly (AIA) on board the Solar Dynamics Observatory (SDO), we studied a limb coronal shock wave and its associated extreme ultraviolet (EUV) wave that occurred on 2010 June 13. Our main findings are (1) the shock wave appeared clearly only in the channels centered at 193 \AA\ and 211 \AA\ as a dome-like enhancement propagating ahead of its associated semi-spherical CME bubble; (2) the density compression of the shock is 1.56 according to radio data and the temperature of the shock is around 2.8 MK;
(3) the shock wave first appeared at 05:38 UT, 2 minutes after the associated flare has started and 1 minute after its associated CME bubble appeared; (4) the top of the dome-like shock wave set out from about 1.23 R$_\sun$ and the thickness of the shocked layer is $\sim$ 2$\times$10$^4$ km; (5) the speed of the shock wave is consistent with a slight decrease from about 600 km s$^{-1}$ to 550 km s$^{-1}$; (6) the lateral expansion of the shock wave suggests a constant speed around 400 km s$^{-1}$, which varies at different heights and directions. Our findings support the view that the coronal shock wave is driven by the CME bubble, and the on-limb EUV wave is consistent with a fast wave or at least includes the fast wave component.
  
 \end{abstract}
\keywords{Sun: eruption, Sun: EIT wave, Sun: EUV emission, Sun: CME, Sun: shock waves, Sun: radio emission}

\section{Introduction}
In 1997, the EUV Imaging Telescope (EIT; \citealt{Delaboudiniere95}) on board the Solar and Heliospheric Observatory (SOHO; \citealt{Domingo95}) observed globally propagating wave-like disturbances in the corona, known as EUV waves (also called ``EIT'' waves), for the first time \citep{Moses97,Thompson98}. Usually, EUV waves are observed as diffuse and irregular arcs of increasing coronal emission in the 195 \AA\ channel. Compared to Moreton waves, EUV waves are a relatively frequent phenomenon and their speeds are relatively low. According to \citet{Biesecker02}, about 7\% of the events in their large sample displayed sharp and bright wavefronts somewhat reminiscent of Moreton waves \citep{Thompson00}. Such EUV waves are also called ``bow waves'' by \citet{Gopalswamy00} or ``S-waves'' by \citet{Biesecker02}. In several events the S-waves coincide spatially with Moreton waves observed at the same time \citep{Khan02,Warmuth04}, which would imply that at least S-waves are the long-sought coronal counterpart to Moreton waves. It is generally accepted that both Moreton waves and S-waves are shock waves.

 However, the physical nature of the more common diffuse EUV waves is still being debated among fast-mode MHD waves \citep[e.g.,][]{Wills99,Wang00,Wu01,Ofman02,Schmidt10}, slow-mode MHD waves \citep{Wang09} and non-waves related to a current shell or successive restructuring of field lines during the coronal mass ejection \citep{Delannee00,Delannee08,Chen02,Chen05,Attrill07}. The observations from the Extreme Ultraviolet Imager (EUVI; \citealt{Wuelser04,Howard08}) instruments on board the twin Solar-Terrestrial Relations Observatory (STEREO, \citealt{Kaiser08}) spacecraft have advanced our understanding of the diffuse EUV wave \citep[e.g.,][]{ Long08,Veronig08,Gopalswamy09,Attrill09,Patsourakos09,Ma09,Kienreich09,Zhukov09,Veronig10,Yang10}. A hybrid wave/non-wave hypothesis was first suggested by \citet{Zhukov04} and has been extended using computational simulation \citep{Cohen09,Downs11}. For detailed discussions of the different models and supporting observations, we refer to the recent reviews by \citet{Warmuth07,Warmuth10,Vrsnak08,Wills09} and \citet{Gallagher10}. 

Due to the scarcity of direct observations of shock waves with high spatial resolution, the well-known signatures $\sbond$ type II radio bursts \citep{Wild50,Wild54,Nelson85} $\sbond$ have long been used to indicate shock waves. More recently, coronagraphic observations in white light \citep{Ontiveros09} and UV spectra \citep{Raymond00,Mancuso09} have also been used to study coronal shock waves. \citet{Klassen00} made a statistical study of type II bursts, and found that 90\% are associated with an EUV wave. However, the exact relationship between the EUV and shock wave is still open because of the low (or even no) spatial resolution of the radio data. For the same reason, the origin of the shock wave is also open. Two interpretations have been proposed, one suggesting a blast wave ignited by the pressure pulse of a flare and the other arguing a piston-driven shock due to a CME \citep[e.g.,][]{Reiner01,Lin06,Oh07,Magdalenic08,Vrsnak08}.  

Recently, \citet{Veronig10} presented a study of a weak dome-like shock wave observed by EUVI which proved the feasibility of using EUV observations to study shock waves. The Atmospheric Imaging Assembly (AIA, \citealt{Title06,Lemen11}) on the Solar Dynamics Observatory (SDO) has 7 EUV and 2 UV wavelengths, covering a wide range of temperatures, at high time cadence (12s) and resolution (1$\arcsec$.4, with 0$\arcsec$.6 pixels). Taking advantage of these capabilities, \citet{Liu10} studied a global coronal EUV wave and found multiple ``ripples'' for the first time; their results support the hybrid EUV wave hypothesis. Direct observations of shock waves with high spatial resolution in the low corona would greatly improve our understanding of the origin of shock waves and the nature of diffuse EUV waves (hereafter EUV wave).

In this paper, we present a detailed analysis of the 2010 June 13 event (cf. \citealt{Kozarev11}), including time-dependent ionization and the nature of EUV wave near the solar surface. We briefly introduce the data used in this paper in the next section. The observations and results are presented in Section 3, and we present our discussion and conclusions in Section 4.

\section{Data and Observation}
On 2010 June 13, a limb solar eruption occurred in the active region NOAA 11079 (S24W91) from around 05:30 UT to 05:50 UT. This event involved a filament eruption, an M1.0 flare and a CME with a strong and short-lived acceleration phase that has been studied in detail by \citet{Patsourakos10}. In particular, a distinct dome-like shock wave associated with this eruption was observed clearly by the AIA imagers in the coronal channels centered at 193 \AA\ and 211 \AA, allowing us to study the morphology and propagation characteristics of the shock wave in detail. In addition, a type II burst was also observed by some radio spectrographs, indicating the presence of a shock wave. 

To study the shock wave and associated phenomena, we mainly used the following data:
\begin{itemize}
\item The AIA on SDO provides multiple simultaneous high-resolution full-disk images of the corona and transition region up to 0.5R$_\sun$  above the solar limb with 1.5 arcsec spatial resolution and 12 s temporal resolution \citep{Lemen11}. Seven narrow EUV bandpasses centered on specific lines: Fe XVIII (94 \AA), Fe VIII,XXI (131 \AA, Fe IX (171 \AA), Fe XII,XXIV (193 \AA), Fe XIV (211 \AA), He II (304 \AA), and Fe XVI (335 \AA) have been employed. The temperature diagnostics of the EUV emissions cover the range from 6$\times$10$^4$ K to 2$\times$10$^7$ K \citep{Lemen11}. Here, we mainly use the channels centered at 171 \AA, 193 \AA, 211\AA\ and 335 \AA\ (Level 1.5 images). To show the shock wave more clearly we employed an improved radial filter technique first developed by S. Cranmer and A. Engell following \citet{Ma10}. The radial filter technique involves dividing the sun into different concentric ``rings'' and calculating the minimum (Imin) and maximum (Imax) intensity in each radial ring. The scaled intensity at each pixel in the ring is then obtained by using the equation I$_{scaled}$= (I-Imin)/(Imax-Imin). 

\item  The data from EUVI onboard the STEREO Ahead spacecraft (STA). STA is 73.6$\degr$ ahead of AIA on its orbit during the eruption, therefore this event is a limb event for AIA and an on-disk event for STA. The two different points of view allow us to construct the 3-D structure of the coronal wave and pursue the nature of coronal waves. EUVI observed the chromosphere and low corona in four spectral channels (304 \AA, 171 \AA, 195 \AA\ and 284 \AA) out to 1.7 R$_\sun$ with a pixel-limited spatial resolution of 1$\arcsec$.6 pixel$^{-1}$ \citep{Wuelser04}. Here we only use the 195 \AA\ images from EUVI A with a time cadence of 2.5 minutes. 
\item The data from the solar radio spectrograph of San Vito observatory, which sweeps the frequency range 25 to 180 MHz every 3 seconds. It monitors solar radio emissions originating mainly in the solar corona. It has a low band (25 to 75 MHz) antenna (non-tracking semi-bicone) and a high band (75 to 180 MHz) antenna (tracking log-periodic). The radio data used in this paper was downloaded from the USAF Radio Solar Telescope Network (RSTN) solar radio fixed frequency and spectral data at the NOAA National Geophysical Data Center (NGDC) Website (\url{http://www.ngdc.noaa.gov}).  
\end{itemize}

\section{Results}
\subsection{Shock Wave in EUV Wavelengths}

\subsubsection{Physical Characteristics}

Figure 1 shows the AIA observations (processed using the radial filter technique) of this event in the 171\AA, 193 \AA\ and 211\AA\ channels. The top row images in Figure 1 show the morphology of the corona before the shock wave appeared and the bottom row images show the morphology during the shock wave propagation. During the eruption, a sharp spherical CME bubble (indicated by the black arrows in Figure 1) can been seen in all three channels. For the first time, a spherical shock wave (indicated by white arrows in Figure 1) was observed by AIA at EUV wavelengths (see the online animations: aia\_20100613\_171.mpg, aia\_20100613\_193.mpg and aia\_20100613\_211.mpg). Note that this event was also which is also studied by \citet{Kozarev11}. The shock wave was ahead of the CME bubble from its first appearance in AIA images at about 1.23 R$_\sun$ (from Sun center) at $\sim$05:38 UT. The thickness of the layer of shocked gas is $\sim$2$\times$10$^4$ km based on the measurement from AIA images along the radial direction (meridional 116 degree) in the 193 \AA\ channel. 

Interestingly, the shock wave appeared as a distinct bright feature only in the 193 \AA\ and 211 \AA\ channels. No clear signature of the shock wave can be identified from the 171 \AA\ image in Figure 1, but the online animation aia\_20100613\_171.mpg shows that the shock wave appeared as a dark feature in the 171 \AA\ channel. The 193 \AA\ and 211 \AA\ channels are dominated by the Fe XII lines [log T$\sim$6.2] and Fe XIV lines [log T$\sim$ 6.3] for AR observations, respectively. The 171 \AA\ channel observes the Fe IX line [log T $\sim$ 5.85] for active region (AR) plasma (see \citealt{O'Dwyer10} for detailed information). We show below that the temperature of the dominant plasma in the shock wave is around 2.8 MK.

The white box in Figure 1 mark the area where we measured the intensity change before and during the passing of the shock wave. The intensity tracking in the white box is placed in Figure 2, with the intensities in 211 \AA\ and 335 \AA\ scaled by multiples 5 and 250, respectively. The plot shows that after the shock wave passed, the intensity in the 193 \AA, 211 \AA\ and 335 \AA\ channels increased rapidly and then went through a rapid decrease followed by a slow one. However, the intensity in 171 \AA\  channel decreased when the shock wave passed and increased slowly until the CME bubble arrived. The intensity jumps (maximum intensity divided by the pre-shock intensity) are 1.2, 1.5 and 1.7 in the 193 \AA, 211 \AA\ and 335 \AA\ band, respectively.

\subsubsection{Upward Propagation}
The three dotted lines in the bottom panels in Figure 1 indicate the radial directions at 115$\degr$, 116$\degr$ and 117 $\degr$ clockwise from the solar north pole, respectively, along which we analyzed the upward propagation of the shock wave and associated CME. 

Figure 3 contains time-distance slit images showing the shock wave formation and propagation in different wavelengths along the three radial directions marked by the dashed black lines in Figure 1. The images in the left, middle and right columns are taken from the 171 \AA, 193 \AA\ and 211 \AA\ channels, respectively. For each column, the top, middle and bottom panels are obtained by placing the slit along the radial directions at 115$\degr$, 116$\degr$ and 117 $\degr$. The dashed curves in each panel of Figure 3 are the GOES flux in 1-8 \AA\ passband showing the associated flare. A filament associated with this eruption rose sharply beginning at 05:32 UT (panel d and g in Figure 3), indicating that the filament initiated this solar eruption.  

The CME bubble is most apparent in the 171 \AA\ channel, so we use it to study the propagation of the CME bubble. In each radial direction, the CME bubble shows similar propagation: going through a rapid acceleration phase and then mpging with a nearly constant or slightly decreasing speed. The slit image at 117$\degr$  in the 171 \AA\ channel (panel c of Figure 3) is then used to measure the distance, speed and acceleration of the CME bubble. The result is given in the top panel of Figure 4 with the red, blue and orange lines with plus symbols indicating the distance, third degree polynomial fit speed and acceleration of the CME, respectively. The upward speed of the CME initially increased from 0 to 500 km s$^{-1}$ and then decreased to 250 km s$^{-1}$ before leaving the AIA field of view. The corresponding acceleration of the CME decreased from around 6 to -4 km s$^{-2}$. (For a detailed analysis of the CME, refer to \citealt{Patsourakos10}.)

The shock wave can be identified in the 193 \AA\ and 211 \AA\ channels, although its appearance differs slightly in each. In the 193 \AA\ channel, the shock wave front is relatively sharp and distinctly separated from the CME bubble (see panels d-f of Figure 3). However, in the 211 \AA\ channel, the intensity in the region between the shock wave front and the CME bubble appears evenly increased (see panels g-h of Figure 3). As discussed in Section 3.3, this is because it takes a finite time to ionize the plasma from Fe XII, which dominates the 193 \AA\ channel, to Fe XIV, which dominates the 211 \AA\ channel. The shock wave first appeared in AIA images at 1.23 R$_\sun$ from the surface of the Sun at around 05:38 UT (dotted-dashed line in panel e of Figure 3), two minutes after the CME showed a distinct rise (dotted line). For measuring convenience, we chose panel e of Figure 3 to analyze the shock wave propagation along the radial direction. Rough estimates showed that the linear fit speed of the shock wave is around 600 km s$^{-1}$, larger than the CME bubble's linear fit speed of 410 km s$^{-1}$ (not including the rapid rising phase).  

We identified the brightest point of the shock wave in each image and used a semi-automatic tracking method to obtain the distance of the shock wave along the radial direction. The result is shown in the top panel of Figure 4, with the red, blue and orange lines with diamond symbols giving the distance, third degree polynomial fit speed and acceleration of the shock wave front. The third polynomial fit speed of the shock wave showed a slight decease from 600 km$^{-1}$ to 550 km s$^{-1}$ with a monotonically decreasing acceleration less than -1 km s$^{-2}$. 

The bottom panel of Figure 4 displays GOES soft X-ray flux. The orange curve for 1-8 \AA, the purple curve for 0.5-4 \AA\ and the green curve is the time derivative of the 1-8 \AA\ flux. The three dotted vertical lines in each of panels from left to right in Figure 4 mark the times: 05:36 UT, 05:37 UT and 05:38 UT, respectively. The Figure shows that the flare began at around 05:35:30 UT and peaked at 05:39 UT. The distinct disturbance of coronal loops which later became part of the CME bubble first appeared at 05:36 UT (left line), while the sharp CME bubble came into being at 05:37 UT (middle line). The shock wave appeared at 05:38 UT (right), which is later than the first appearance of both flare and CME. 
 
\subsubsection{Lateral Propagation and EUV Wave}
Figure 5 gives base difference images showing the evolution of the shock wave and the associated EUV wave. The EUV wave here refers to the disturbance propagating laterally along the solar surface which normally appears as a projected bright circular feature propagating on the solar disk in EUV difference images. Panels a-h of Figure 5 are AIA base difference images in 193 \AA\ channel obtained by subtracting the 193 \AA\ image at 05:32:08 UT from the present images. Panels i-l of Figure 5 are EUVI A images in the 195 \AA\ channel obtained by subtracting the 195 \AA\ image at 05:33:00 UT from the current images. The AIA data (panels a-h, see also the online animation basedif\_aia\_20100613.mpg) show a close relationship between the shock wave and the EUV wave: 1) they appeared at the same time $\sim$ 05:38:08 UT; 2) the front of the laterally expanding shock wave is tightly connected to the position of the EUV wave front. In EUVI observations, we could not identify the coronal shock wave, probably because the shock wave is weak and the column depth is small. The EUV wave can be identified in EUVI as a circular disturbance propagating from AR 11079 (see also the online animation EUVI\_A\_20100613.mpg). 

The top panel of Figure 6 is a projected base difference image (see the online animation project\_diff\_20100613\_193.mpg for more information) where the horizontal coordinate is polar angle (clockwise from north pole of the Sun) and the vertical coordinate is radial distance in solar radii. The five horizontal lines (which are parallel to the solar surface) from bottom to top indicate the layers located at 0.98, 1.01, 1.04, 1.07 and 1.11 R$_\sun$, respectively. By placing the slit along these layers, we obtained five different slit images from the base difference images in 193 \AA. These slit images are placed in panels b-f of Figure 6. The horizontal coordinate of the slit images is the meridional angle in degrees while vertical coordinate is time in minutes. The black area indicates the dimming area changing with time and the white features indicate the EUV brightening. Taking the propagating white feature as the coronal wave front, we estimate the EUV wave speed. The dotted lines and the numbers around them in Figure 6 indicate the linear fit distances of the EUV wave fronts and the estimated speeds. The propagation of the EUV wave towards the north (left) and south (right) is not symmetric; the speed in the southward direction appears slower than the northward direction.

A bright feature was also observed propagating in the reverse direction, implying the reflection of the EUV wave (indicated by the ``RF'' in panel d and e of Figure 6). However, a detailed analysis of this observation is left for future work.
   
\subsection{Shock Wave in Radio Observation}

According to the National Geophysical Data Center¡¯s event listing\footnote[1]{ftp://ftp.ngdc.noaa.gov/STP/SOLAR\_DATA/SOLAR\_RADIO/SPECTRAL/2010/SPEC\_NEW.10}, a type II burst associated with this event was observed by several radio spectrographs. In Table 1 we list the extracted dates, observatory stations, event start time, event end time, spectral class (SC), lower frequency (LF), upper frequency (UF) and estimated shock speeds in km s$^{-1}$. The start time of the type II fits well with the AIA observation of the shock wave and there is no other solar eruption at the same time. Therefore, the shock wave observed at radio and EUV wavelengths should be the same one. 
\begin{table}[ht]
\caption{Record of the type II occurred on June 13, 2010.}
 \begin{tabular}{c c c c c c c c }
  \hline
 Station& Start time & End time & SC & LF(MHz)& UF(MHz) & Shock Speed (km/s)\\
 \hline
 BLEN     &   0537.0  & 0541.1      &II     &   175X  &  332    &  \\
 CULG    &   0537.0   & 0543.0     &II     &    57X   & 150     &   \\
 HIRA     &   0537.0   &  0550.0     &II    &    50     &  310    &   \\
 CULG    &  0538.0    & 0551.0      &II    &   57X   & 200     & 700\\
 SVTO    &  0538.0    & 0552.0      & II   &   35     & 180     & 621\\
 LEAR     &  0539.0    & 0548.0      & II   &   55      & 180    & 665\\
\hline
\end{tabular}
\end{table}

Figure 7 shows the metric type II radio emissions from San Vito radio (SVTO) spectrograph. The top panel is the dynamic spectrogram. The type II burst occurring from 05:38 UT to 05:53 UT can be easy identified.  The black curve indicates the harmonic frequency (2$f_p$) in the type II burst along which we measured the local plasma frequency ($f_p$) drift. The result showed that 2$f_p$ drifts from 165 MHz to 40 MHz. Considering the relationship between plasma frequency and density 
 \begin{equation}
 f_{p}=8.98\times10^{3}\sqrt{n},
  \end{equation}
we determine the density of the shock wave front (middle panel of Figure 7). The density of the shock wave dropped from $8.6\times10^{7}$ cm$^{-3}$ at around 05:38 UT to $4.5\times10^{6} $cm$^{-3}$ at about 05:52 UT. The inset in the middle panel is the coronal plasma density model from \citet{Sittler99} as used by \citet{Lin06}, 
\begin{equation} n(z)=n_0\,a_1\,z^{2}\,e^{a_2\,z}[1+a_{3}\,z+a_{4}\,z^{2}+a_{5}\,z^{3}]
\end{equation}
\begin{equation*}
z=1/(1+y), a_1=0.001272, a_2=4.8039,
\end{equation*}
\begin{equation*}
a_3=0.29696, a_4=-7.1743,a_5=12.321
\end{equation*}
where y is the height above the solar surface in solar radii and $n_0$ is the electron number density at the solar surface, which is chosen as $10^9 $ cm$^{-3}$ here. The density model was used to estimate the position of shock wave front (the black line in the bottom of Figure 7), showing that  the shock wave formed at around 1.25 R$_\sun$.  The derived speeds are displayed by the dotted and dashed lines in the bottom panel of Figure 7. The dotted line shows the linear fit speed (527 km s$^{-1}$) and the dashed line shows the second degree polynomial fit speed which decreased from around 600 km s$^{-1}$ to 400 km s$^{-1}$. The speed of the shock wave estimated using the model of \citet{Sittler99} is smaller than the speeds listed in Table 1. As the speed derived from radio emissions is strongly dependent on the coronal density model used, the error of the speed is quite large. Considering this situation, the speed of the shock wave measured from radio emission is consistent with that obtained from AIA observation.   
  
The two black stars mark the frequencies used to measure the density jump at the shock using the equation,  \begin{equation} X= \frac {n_2}{n_1}=(\frac {f_U}{f_L} )^2.
\end{equation}
As described in \citet{Vrsnak02}, the band-split frequencies in type II emission map the electron densities behind and ahead of the shock front. In front of the shock the plasma is characterized by the electron density $n_1$ and emits radio waves at the frequency $f_L$ (lower frequency branch) while the plasma behind the shock is characterized by the electron density $n_2$ and emits radio waves at the frequency $f_U$.  At 05:40 UT, the lower and upper frequencies of the harmonic bands in type II burst are around 132$\pm$5 MHz and 165$\pm$15 MHz (marked by the black stars in the top panel of Figure 7), respectively. Using equation (3), we obtained the density jump $X=1.56\pm0.1$ at 05:40 UT. \citet{Gopalswamy11} obtained a similar result using radio data from HiRAS. For an oblique shock (taking the adiabatic index $\gamma=5/3$) the Alfv\'{e}n Mach number $M_A$ and the density jump $X$ are related by: 
 \begin{eqnarray}
(M_{A}^2-X)^2[5\beta X+2M_{A}^2\cos^{2}\theta(X-4)]   \nonumber   \\
 + M_{A}^2\,X\sin^{2}\theta[(5+X)M_{A}^2+2X(X-4)]=0
  \end{eqnarray}
\citep{Vrsnak02}. For $\beta$ $\rightarrow$ 0, in the case of the perpendicular shock ($\theta=90\degr$),  
$M_{A}$=$\sqrt{X(X+5+5\beta)/2(4-X)}$=1.45 and in the case of longitudinal shock ($\theta$=0$\degr$), $M_{A}$=$\sqrt{X}$=1.25. Considering $M_{A}$=$v/v_A$ and $M_{A}$= 1.35$\pm0.1$ (the actual error maybe larger) this produces an estimated Alfv\'{e}n speed 450$\pm$30 km s$^{-1}$. The estimated sound speed $C_s=\sqrt{(\gamma \kappa T)/m}$ is 156$\pm$30 km s$^{-1}$ for a pre-shock temperature of (1.8$\pm$0.4)$\times$10$^6$ K. According to the DEM solutions for regions 2 and 3 in \citet{Kozarev11}, the pre-shock temperature (peak value) is around $logT= 6.25\pm0.1$ (1.8$\pm$0.4 MK).  The area focused on in this paper differs slightly from the region 2 and 3 studied by  \citet{Kozarev11}, so there may be a small discrepancy in the temperature, although it should not be too different. In the following context, we using 1.8$\pm$0.4 MK as the pre-shock temperature T$_1$. The resulting fast magnetosonic speed $c_f=\sqrt{v_A^2+c_s^2}$ is 476$\pm$38 km $s^{-1}$. In the solar atmosphere with $\tilde{\mu}$=0.6 and $\gamma$=5/3,  the magnetic field strength can be derived (from the equation 2.48b of \citealt{Priest82a}) as $B=3.57\times10^{-4}\, n_0^{1/2}\, V_A$. For a pre-shock density of 6$\times10^7$ cm$^{-3}$ from the type II frequency at around 05:40 UT, the value of $v_A$ implies B $\sim$ 1.3 Gauss. \citet{Gopalswamy11} obtained an estimate for B that is consistent with this using a completely different technique based on the standoff distance between the flux rope and the shock.

\subsection{Verifying the Shock Wave Interpretation}
Coronal shocks have been identified from white light coronagraph images by morphology and density enhancement \citep[e.g.,][]{Vourlidas03,Ontiveros09} and from UV spectra \citep{Raymond00,Mancuso09} by measuring oxygen kinetic temperatures. AIA shows brightening ahead of the bright CME loops, but that could in principle be part of ejected streamer rather than a shock. The typical attributes of a shock are density compression, gas heating and subsequent gas ionization. The intensities obtained from AIA imagers combine the electron density and ionization state of the plasma. 
To test the relation between compression and heating by an MHD shock, we employ V$_s$=600 km s$^{-1}$ from observed speed of shocked front and $\rho_2$/$\rho_1$=1.56 from the band splitting in the radio. If the shock was purely gas dynamic, with a sound speed C$_s$=156 km s$^{-1}$, and a Mach number M=600/156=3.8, the compression would be $\rho_2$/$\rho_1$=3.3, 
instead of the observed 1.56. Therefore the magnetic field B is important. According to the jump condition for a perpendicular shock (Equation 2.19 of \citealt{Draine93}), the compression ratio is 
\begin{equation} \frac{\rho_2}{\rho_1}=\frac{2(\gamma+1)}{D+[D^2+4(\gamma+1)(2-\gamma)M^{-2}_A]^{1/2}}, \end{equation}
where $D=(\gamma-1)+(2M^{-2}+\gamma M_{A}^{-2})$.
Assuming M=3.8 and ${\rho_2}/{\rho_1}$=1.56, we find M$_A$=1.55.

For a perpendicular shock, the shocked plasma parameters ($v_2$, $\rho_2$, $p_2$, $B_2$) are related to those of the unshocked plasma ($v_1$, $\rho_1$, $p_1$, $B_1$) by the equations for conservation of momentum and energy,
\begin{equation} 
p_2+B_2^2/(2\mu)+\rho_2v_2^2=p_1+B_1^2/(2\mu)+\rho_1v_1^2,
\end{equation}
\citep{Priest82a}. Considering  $\rho_2/\rho_1=X$, $B_2/B1=X$, $v_2/v_1=1/X$, $p=nkT$, $c_s^2=\gamma p1/\rho1$, $v_A^2=B_1^2/\mu\rho_1$, $M=v1/c_s$ and $M_A=v1/v_A$,
the relationship between the pre-shock temperature and post-shock temperature can be obtained as
\begin{equation}
\frac{T_2}{T_1}=\frac{1}{X}[1+(1-\frac{1}{X}-\frac{X^2-1}{2M_A^2}) \gamma M^2].
\end{equation}  
Taking X=1.56, M=3.8, M$_A$=1.55 and $\gamma=5/3$,  $T_2$/$T_1$ is around 1.57 according equation (7).  Assuming the pre-shock temperature of $T_1$=(1.8$\pm$0.4)$\times$10$^6$ K as we explained in Section 3.2, the post-shock temperature $T_2$ is around (2.8$\pm$0.6)$\times$10$^6$ K.

Using the ionization rates $q_i$ at T=2.8$\times$10$^6$ K, we can estimate the ionization time scales and then compare them with observed time scales of the band ratios. The ionization time scale to reach Fe XIV (211 \AA) and Fe XVI (335 \AA) can be obtained from the following equations:
 \begin{equation} t_{211}=\frac{1}{n_e}(\frac{1}{q_{FeXII}}+\frac{1}{q_{FeXIII}})\end{equation}
 \begin{equation}t_{335}=\frac{1}{n_e}(\frac{1}{q_{FeXII}}+\frac{1}{q_{FeXIII}}+\frac{1}{q_{FeXIV}}+\frac{1}{q_{FeV}}).
 \end{equation}
Where $q_{FeXII}$, $q_{FeXIII}$, $q_{FeXIV}$ and $q_{FeXV}$ are ionization rate coefficients in units of cm$^3$ s$^{-1}$. Given the temperature of plasma the ionization rate coefficients can be easily obtained from CHIANTI by using the code ioniz\_rate.pro. In Table 2 we list the ionization rate coefficients at 2.4 MK, 2.6 MK, 2.8 MK, 3.0 MK, 3.2 MK, respectively, in unit of cm$^3$ s$^{-1}$. Taking n$_e$=6$\times$10$^7$$\times$1.56=9.4$\times$10$^7$ cm$^{-3}$, the ionization time scales can be obtained by using the equations (8) and (9). The corresponding results are placed in the right two columns in Table 2.
\begin{table}[ht]
\caption{The ionization rate coefficients (in unit of cm$^3$ s$^{-1}$) at different given temperatures and the corresponding estimated ionization time scales.}
\begin{tabular}{lcccccc}
\hline
T (MK) & $q_{FeXII}$& $q_{FeXIII}$&  $q_{FeXIV}$ & $q_{FeXV}$ & t$_{211} (s)$ & t$_{335}(s)$\\
\hline
2.4  & 4.0982035e-10  & 2.4158748e-10  &    1.4844586e-10   &   1.0106319e-10   & 110   & 386\\
2.6  & 4.6829123e-10  & 2.8404318e-10  &    1.7771562e-10   &   1.2294273e-10   & 94   & 323\\
2.8  & 5.2553271e-10  & 3.2686128e-10  &    2.0770653e-10   &   1.4567346e-10   & 83   & 277\\
3.0  & 5.8123741e-10  & 3.6961965e-10  &    2.3806008e-10   &   1.6894433e-10   & 74   & 242\\
3.2  & 6.3520065e-10  & 4.1199241e-10  &    2.6848624e-10   &   1.9249588e-10   & 66   & 215\\
\hline
\end{tabular}
\end{table}

Figure 8 displays time intensity ratio profiles, which are derived from the intensity changes in different wavelengths (Figure 2) for the region marked by the white box in Figure 1. The rise phase of the intensity ratio profile is indicating the ionization time. The ionization time scales of 211 \AA\ and 335 \AA\ obtained from observation are around 100$\pm12$ seconds (marked by the top two vertical lines  in Figure 8) and 275$\pm12$ seconds (marked by the bottom two vertical lines in Figure 8), respectively. In general, the ionization times obtained from observation are consistent with that derived from the theory of shock wave, 85$\pm$13 seconds for t$_{211}$ and 288$\pm$53 seconds for t$_{335}$. In other words, the brightening ahead of the CME bubble is consistent with the shock jump conditions, while it would be purely coincidental if the structure were not a shock. The theoretical estimates of the ionization times are limited by the uncertainty in the pre-shock temperature, and this might be improved by further analysis of the AIA data. We also note that the intensity ratios of the raw band intensities can be misleading, because they convolve the ionization times with the time scale for the increase of emission from the shocked gas compared to the intensity from the foreground and background plasma outside the shock.  We have estimated the foreground and background emission based on the variation of the 171 A band, and in this case the time scales for the intensity ratios are similar to those obtained from the raw band intensities. Moreover, a more detailed analysis would include the pre-CME density profile and the adiabatic cooling as the shocked gas expands. It should be possible to obtain accurate values of the electron temperature at different positions around the shock to study electron-ion temperature equilibration and energy losses to particle acceleration.

\section{Discussion and Conclusions}

On June 13 2010, a CME-associated shock wave was directly observed by AIA. Combining these observations, EUV data from STEREO A and radio data from SRS, we present a detailed study of the shock wave and its low coronal reactions. 
\subsection{Main Findings}
Our main findings are as follows:
\begin{itemize}

\item The shock wave front first appeared at 05:38 UT as a dome-like bright feature, which is clearly distinguishable from its associated CME bubble. The top of the dome-like shock wave originated from about 1.23 R$_\sun$ with the thickness of the shocked layer around 2$\times$10$^4$ km. A type II radio burst occurred at the same time, further confirming the presence of the shock wave. The electron density jump X (the ratio of the plasma density behind the shock and that in front of the shock) is around 1.56 indicating that the shock wave is a weak shock wave. The pre-shock density and the density jump from the radio data, along with a shock speed of 600 km s$^{-1}$, imply a post-shock temperature of 2.8 MK. The ionization time scales of 211 \AA\ and 335 \AA\ obtained from AIA observation are consistent with the ionization time scale derived from shock wave jump relation in general, which proves the dome-like bright feature is a shock wave.

\item  The shock wave appeared distinctly only in the channels centered at 193 \AA\ and 211 \AA\ with the estimated relative intensity of the shock wave front increasing by 1.2, 1.5 and 1.7 in 193 \AA, 211 \AA\ and 335 \AA, respectively. All of the EUVI channels showed an increasing intensity with the exception of the 171 \AA\ channel which showed a clear intensity decrease. 

\item  The upward speed of the shock wave shows a slight decrease from about 600 km s$^{-1}$ to 550 km s$^{-1}$ with a deceleration less than 1 km s$^{-2}$.
The lateral speed of the shock wave shows no acceleration, but varies according to both height and direction. The linear fit lateral speeds towards the north range from 246 km s$^{-1}$ to 397 km s$^{-1}$ while the speeds towards the south range from 342 km s$^{-1}$ to 486 km s$^{-1}$. The lateral speed of the shock wave seems to increase with the height below 1.1 R$_\sun$. In general, the lateral speed of the shock wave is less than its upward speed.  
 \end{itemize}

\subsection{Origin of the Shock Wave} The high time cadence of AIA observations makes the pursuit of the origin of the low coronal shock waves possible. AIA data shows that a filament rising was observed at 05:32 UT, at the very beginning of this solar eruption. About four minutes after the filament rising, a flare and CME bubble appeared nearly simultaneously. At 05:38 UT, two minutes later after the flare and CME bubble appeared, the shocked front came into being.  Considering the time sequence, it is likely that the filament is a trigger for the whole eruption. Because the flare and CME bubble appeared simultaneously and earlier than the shock wave, neither of them can be eliminated as the driver of the shock wave according to the time sequence alone. However, considering the strong similarity of the shock wave and the CME in both morphology (dome-like) and kinematics (slightly decreasing), we would suggest that the CME bubble played the role of a piston in driving the shock wave. 

\subsection{Nature of the EUV Wave} The direct observation of the low coronal shock wave by AIA strongly supports the existence of a dome-like fast mode wave observed in EUV wavelengths. The fastest front of the EUV wave coincides with lateral expansion of the shock wave. Therefore our result supports the hypothesis that the EUV wave is a fast wave or at least includes a fast wave component. Considering the projection effect of the on-disk observation as discussed in \citet{Ma09}, the on-disk observation may include non-wave component. However due to the weakness of on-disk wave front observed by EUVI, we did not get strong evidence for non-wave component as proposed by \citet{Liu10}. More recently, \citet{Chen11} showed an on-disk event and found both fast and slow propagation fronts. However, there are still inconsistencies between the observed phenomena and the different theories proposed to explain the EUV wave. The high cadence observations from SDO should allow these issues to be resolved. 
\acknowledgments
We sincerely thank the anonymous referee for very helpful and constructive comments that improved this paper. We acknowledge the AIA team for the easy access to calibrated data. We are grateful to San Vito Observatory, NOAA's NGDC and RSNT for providing the radio data and corresponding analysis software. The STEREO/SECCHI data are produced by an international consortium: NRL, LMSAL, NASA, GSFC (USA); RAL (UK); MPS (Germany); CSL (Belgium); and IOTA, IAS (France). SM thanks Kelly Korreck, Mark Webb, Ed Deluca and Meredith Wills for helpful discussions. SM is also grateful to Steven Cranmer, Alexander Engell and Henry Winter for help with techniques. SM gratefully acknowledges NASA grant NNX08BA97G, SP02H1701R and NNM07AB 07C. 

\bibliographystyle{apj}
\bibliography{ms_100613.bib} 

\clearpage

\begin{figure}
\plotone{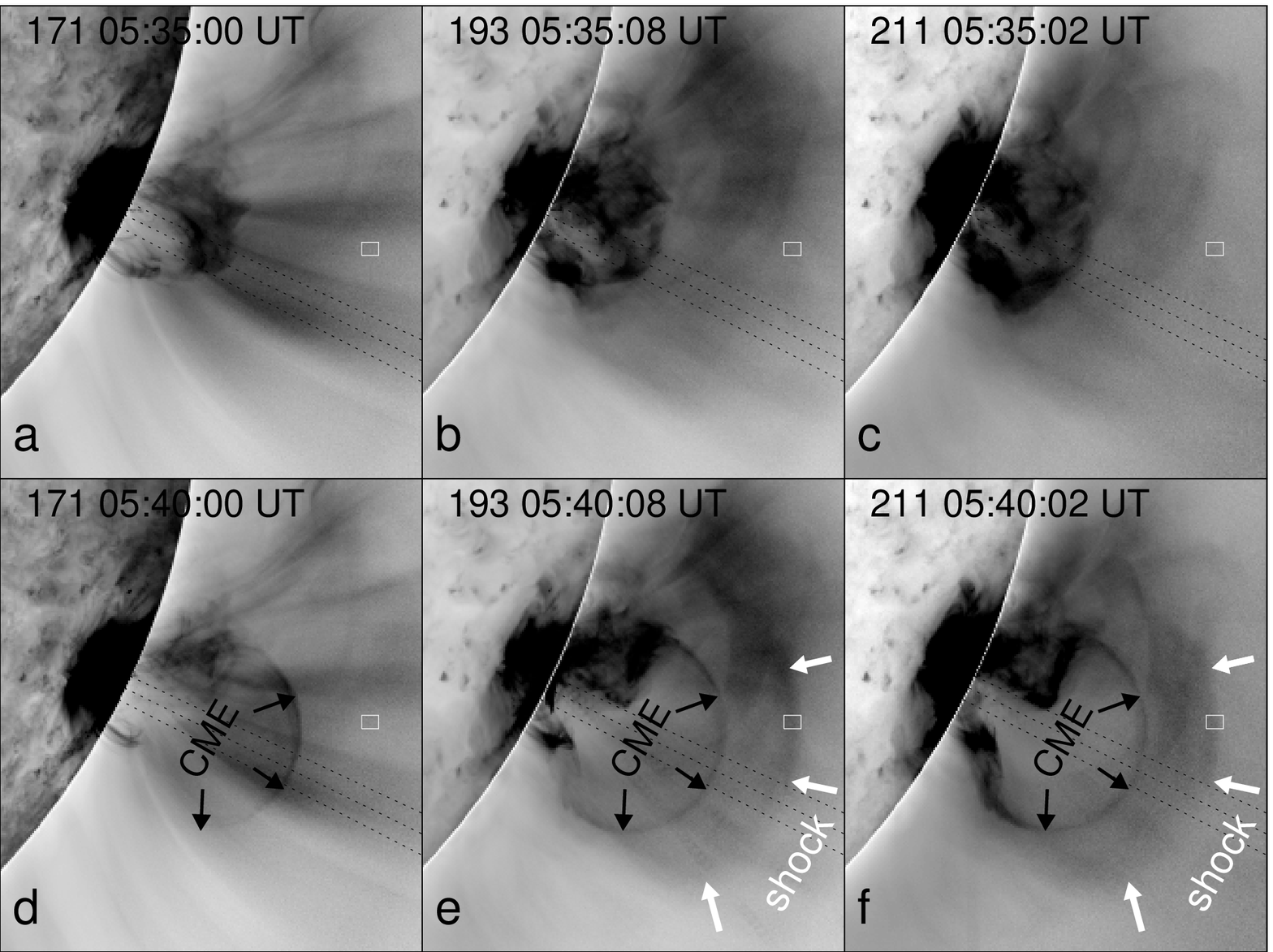} \caption{AIA images showing the morphology of the shock wave (reversed color table). The AIA images were contrast enhanced using a radial filter technique described in Section 2. (Animations aia\_20100613\_171.mpg, aia\_20100613\_193.mpg and aia\_20100613\_211.mpg are available in the online journal.) \label{fig1}}
\end{figure}

\begin{figure}
\plotone{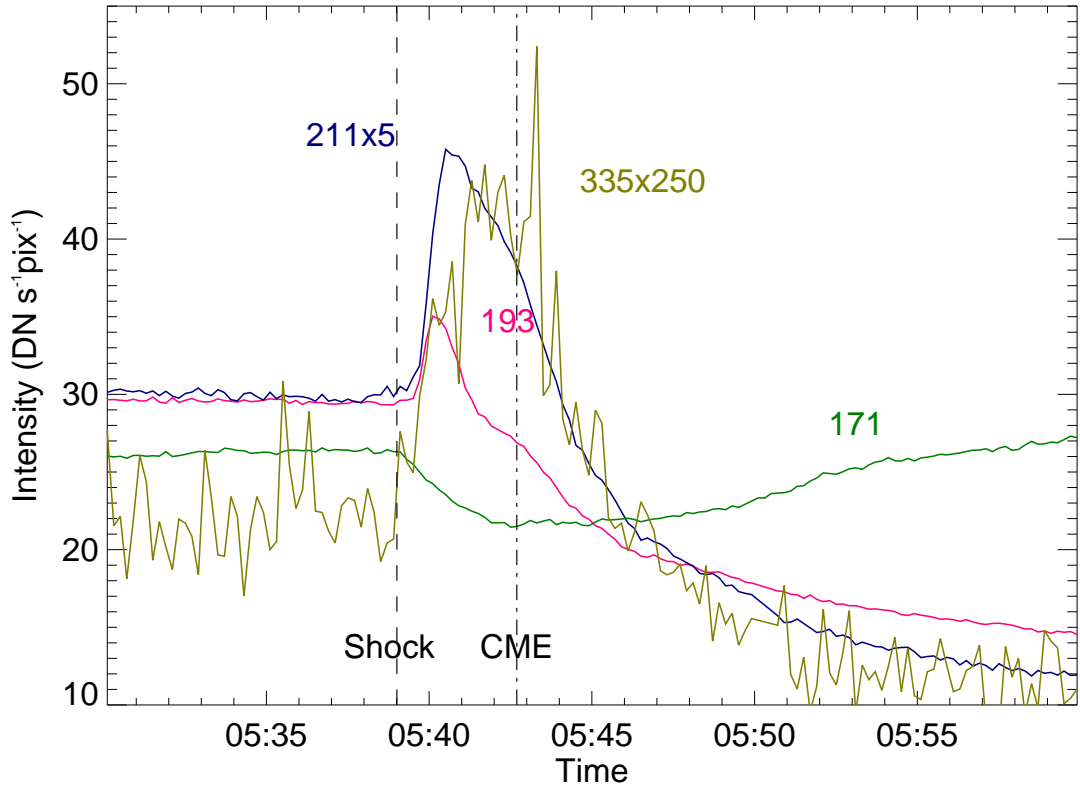} \caption{Intensity flux tracking in the white box in Figure 1. The green, pink, blue and olive curves refer to the intensity flux in 171 \AA, 193 \AA, 211 \AA\ (x5) and 335 \AA\ (x250). The dashed black line and dash-dot line refer to the time when the shock and CME bubble arrived in the white box. \label{fig2}}
\end{figure}


\begin{figure}[hb]
\centering
\includegraphics[scale=0.7,angle=0]{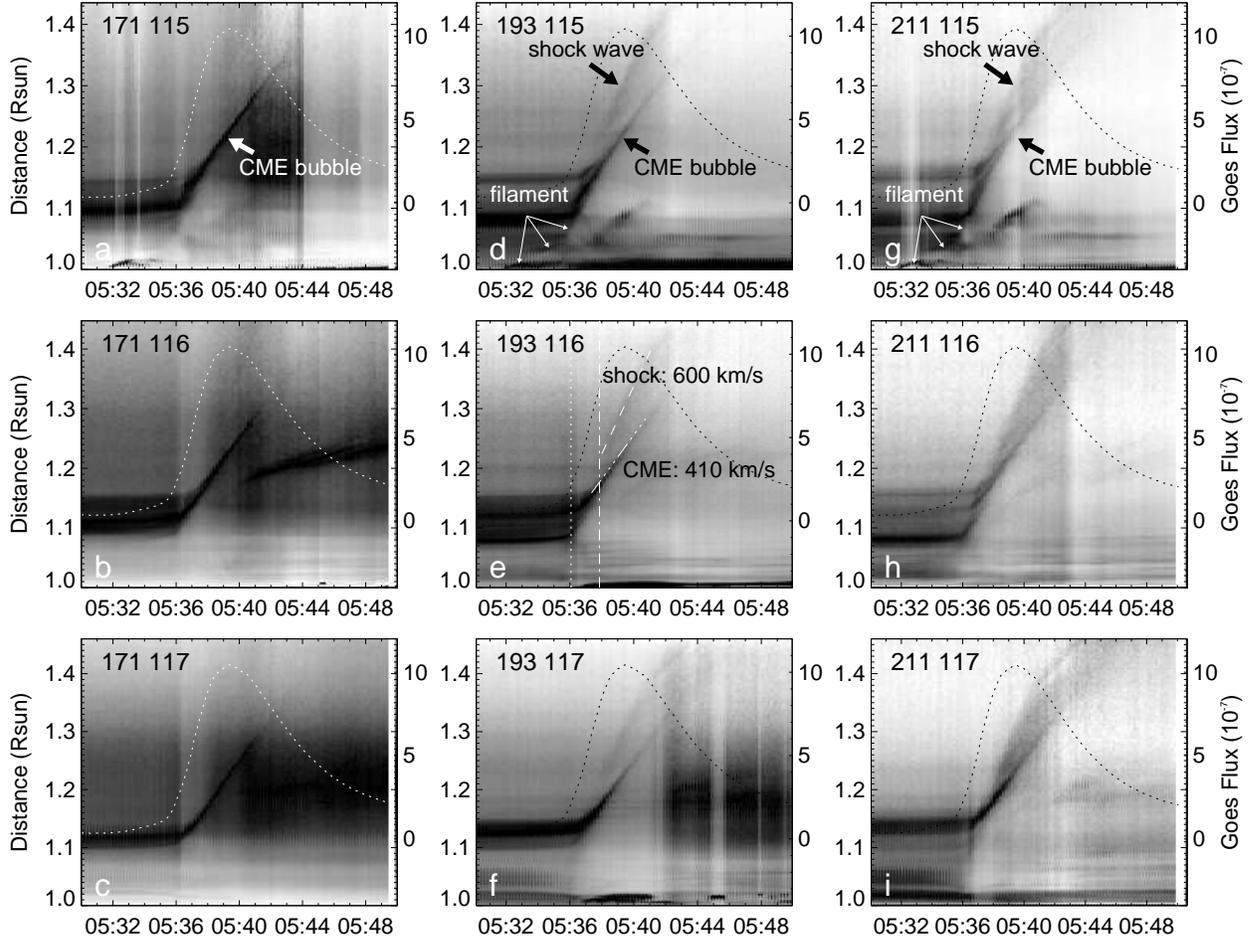} 
\caption{AIA slit images (reversed color table) in 171\AA\ (left column), 193 \AA\ (middle column) and 211 \AA\ (right column) showing the time-distances of the CME bubble and shock wave along 115$\degr$ (top row), 116$\degr$ (middle row) and 117$\degr$ (bottom row), respectively. The contoured dash curves are the GOES 1-8 \AA\ flux. \label{fig3}}  
\end{figure}

\begin{figure}
\epsscale{0.8}
\plotone{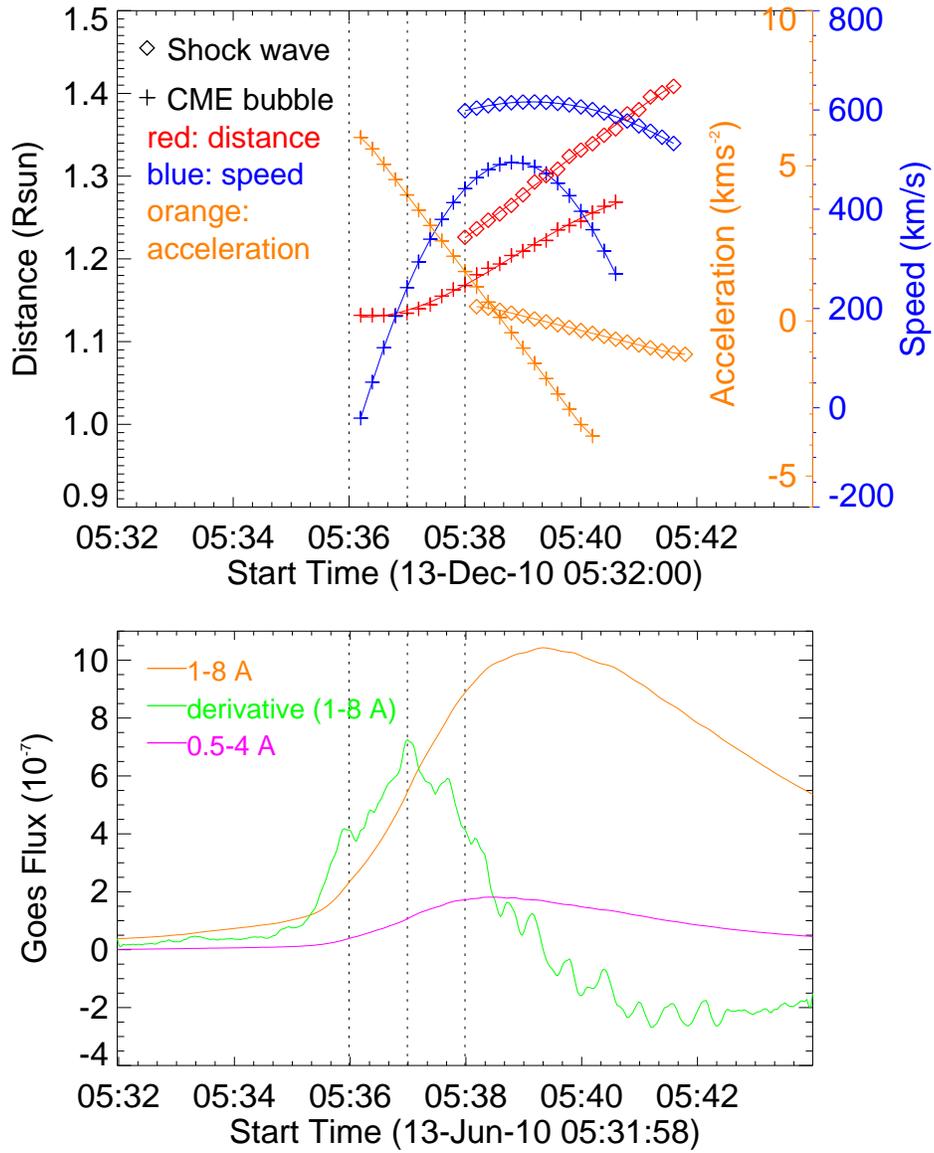} \caption{CME dynamics and GOES flux. The red, blue and orange curves with plus (diamond) symbols indicate the time-distance, time-speed, and time-acceleration profiles of CME bubble (shock). \label{fig4}}
\end{figure}

 \begin{figure}
\plotone{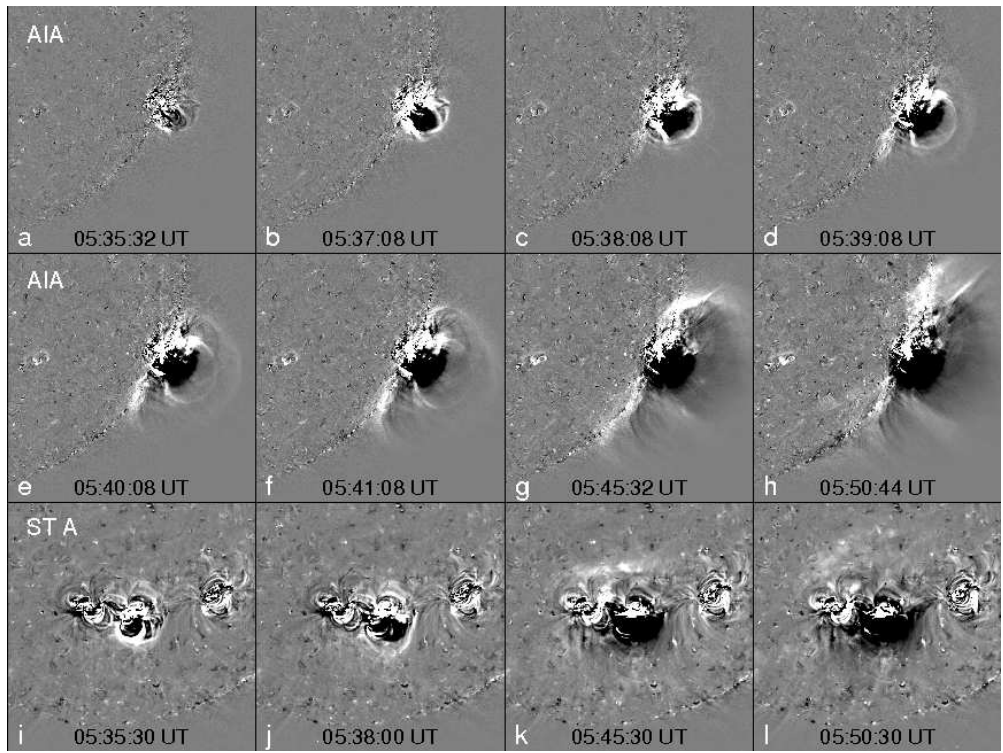} \caption{AIA 193 \AA\ (top two rows) and EUVI A 195 \AA\ (bottom row) observations show the propagation of shock wave and EUV wave. Animations (basedif\_aia\_20100613\_193.mpg and EUVI\_A\_20100613.mpg) are available on line. \label{fig5}}
\end{figure}

\begin{figure}
\epsscale{0.8}
\plotone{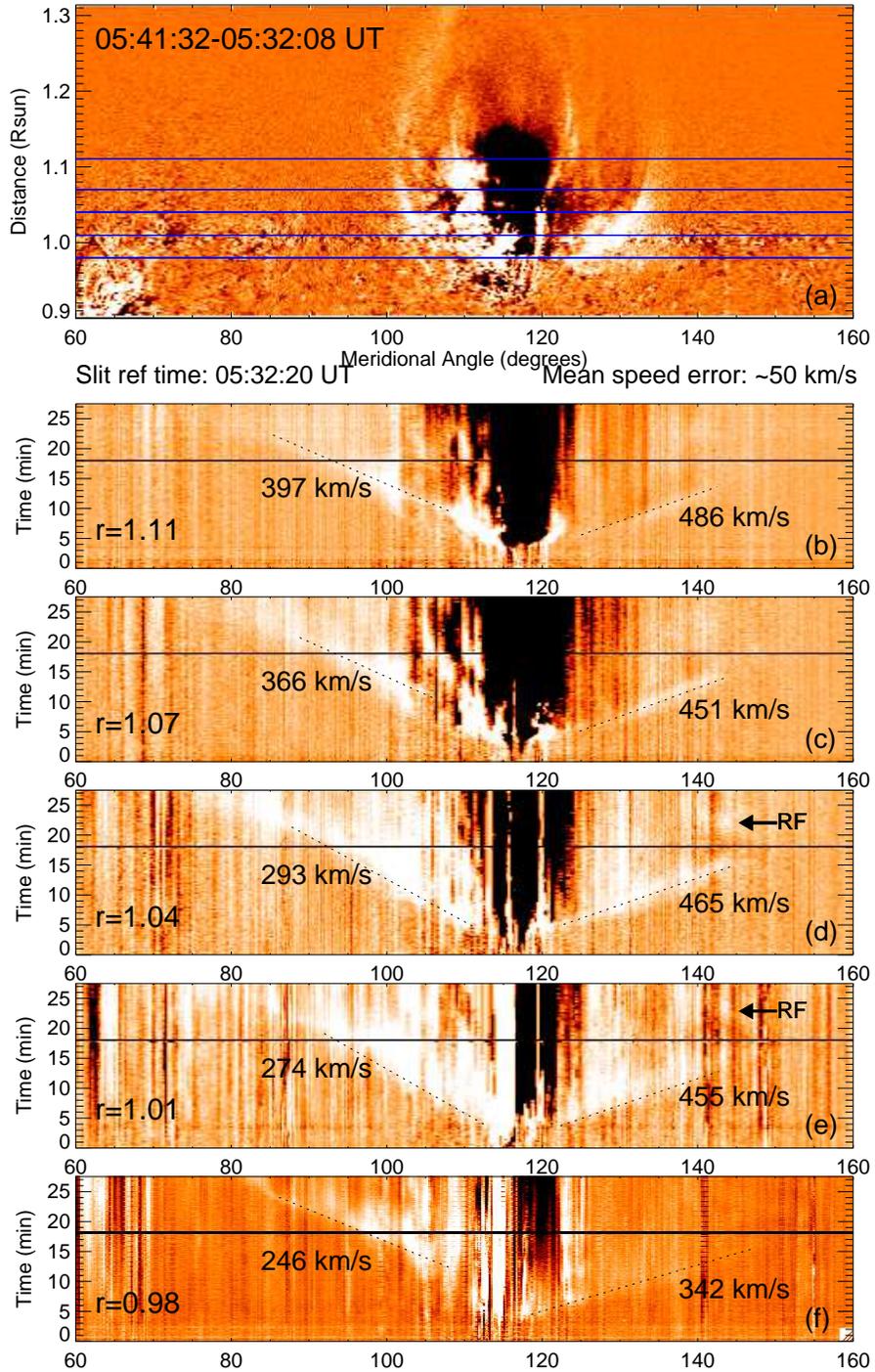} \caption{AIA 193 \AA\ slit images showing the coronal wave propagating at different heights parallel to the solar surface. The base time of the event is 05:32:08 UT. Animation (project\_diff\_20100613\_193.mpg) is available on line. \label{fig6}}
\end{figure}

\begin{figure}
\epsscale{0.8}
\plotone{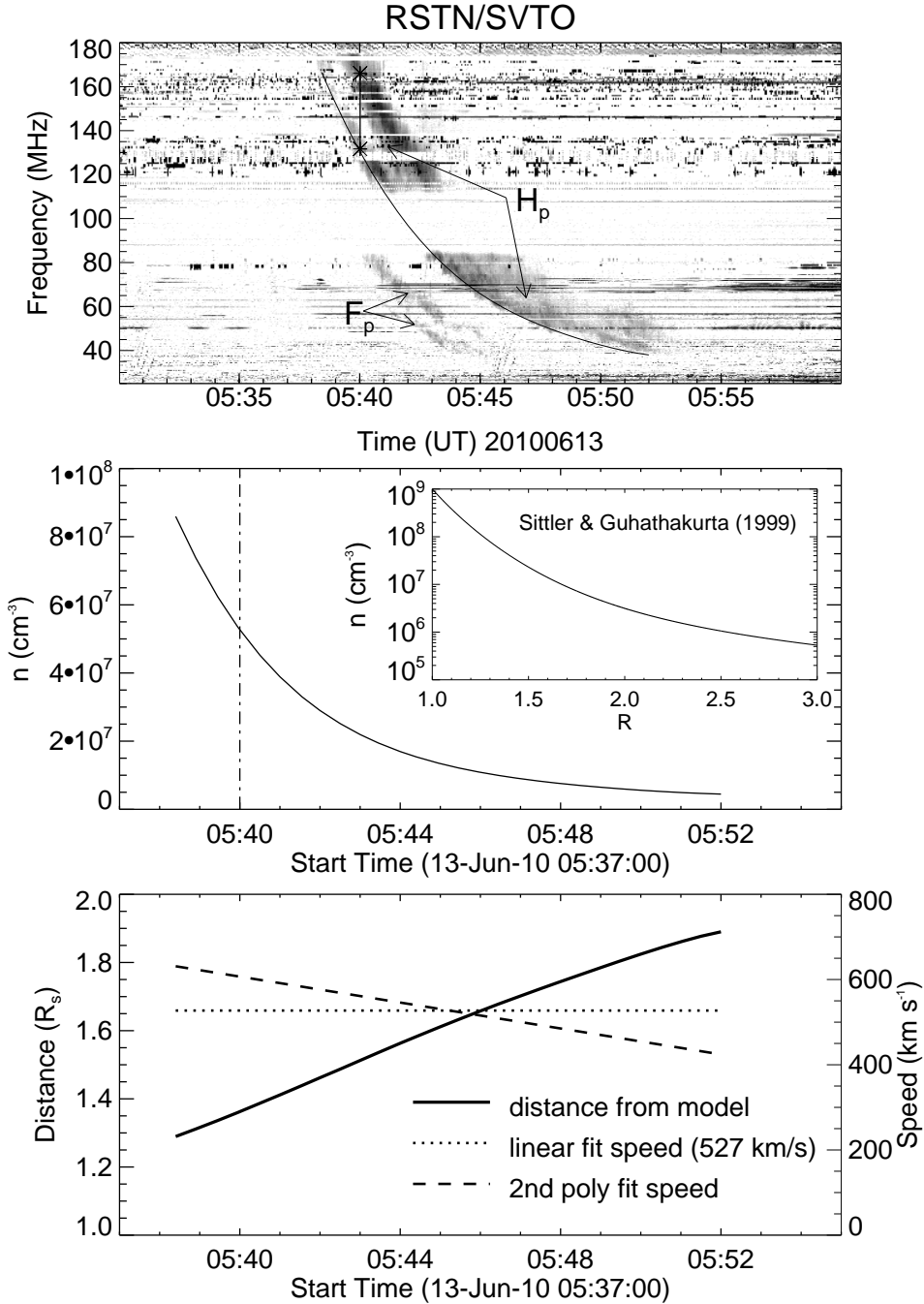} \caption{The dynamic spectrogram from San Vito observatory in the range 25-180 MHz and the derived density and speed of the shock wave. \label{fig7}}
\end{figure}

\begin{figure}
\epsscale{1}
\plotone{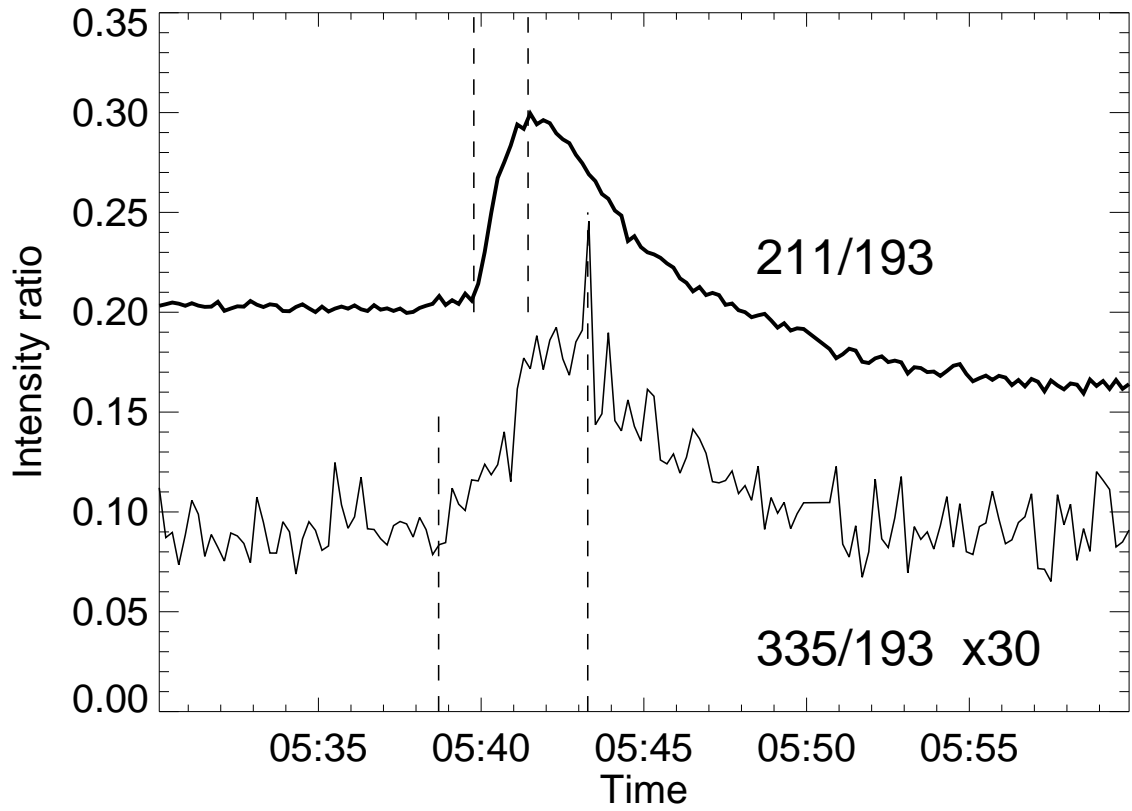} \caption{Time intensity ratio profiles from the white box in Figure 1. The thick black curve for the intensity ratio of 211/193 and the thin black curve for the intensity ratio of 335/193 (x30). \label{fig8}}
\end{figure}

\clearpage

\end{document}